\pdfoutput=1
\documentclass[11pt]{article}

\usepackage[final]{acl}
\usepackage{amsmath}
\usepackage{multirow}
\usepackage{booktabs}
\usepackage{adjustbox}

\usepackage{times}
\usepackage{latexsym}

\usepackage[T1]{fontenc}
\usepackage[utf8]{inputenc}

\usepackage{microtype}

\usepackage{inconsolata}
\usepackage{graphicx}

\title{Cohort Retrieval using Dense Passage Retrieval}

\author{Pranav Jadhav \\
  \texttt{pranavjadhav@tricog.com}}

\begin{document}
\maketitle
\begin{abstract}
Patient cohort retrieval is a pivotal task in medical research and clinical practice, enabling the identification of specific patient groups from extensive electronic health records (EHRs). In this work, we address the challenge of cohort retrieval in the echocardiography domain by applying Dense Passage Retrieval (DPR), a prominent methodology in semantic search. We propose a systematic approach to transform an echocardiographic EHR dataset of unstructured nature into a Query-Passage dataset, framing the problem as a Cohort Retrieval task. Additionally, we design and implement evaluation metrics inspired by real-world clinical scenarios to rigorously test the models across diverse retrieval tasks. Furthermore, we present a custom-trained DPR embedding model that demonstrates superior performance compared to traditional and off-the-shelf SOTA methods. To our knowledge, this is the first work to apply DPR for patient cohort retrieval in the echocardiography domain, establishing a framework that can be adapted to other medical domains.

\end{abstract}

\section{Introduction}

The problem of cohort retrieval has been a cornerstone of medical research and clinical practice since the digitization of medical records. 
In medical research, a cohort refers to a group of patients sharing common characteristics or experiences within a defined period and cohort retrieval is the extraction of specific patient groups or cohorts from extensive medical databases based on predefined criteria. With the growing volume of EHRs, consisting of both structured and unstructured clinical reports, retrieving specific patient cohorts has become a critical task with far-reaching implications. From enrolling eligible patients in clinical trials based on their medical history to deriving demographic-based insights for targeting specific conditions, cohort retrieval drives advancements in healthcare delivery. It is also used in carrying out longitudinal studies, allowing researchers to track patient health trends over years, offering insights into disease progression and treatment efficacy.

To focus the scope of this work, we concentrate on Echocardiography \citep{Kossaify2014EchocardiographyPractice}, a specialized domain within the biomedical field. Echocardiography leverages ultrasound technology to provide real-time visualization of the heart’s structure and function. This non-invasive imaging technique plays a crucial role in diagnosing and managing cardiovascular conditions, offering insights into cardiac anatomy, valve functionality, and blood flow dynamics. However, while integrating echocardiographic data into EHRs has advanced patient care, it has also introduced challenges in data retrieval and analysis. The complex, disorderly nature, and the sheer volume of echocardiographic reports demand advanced Information Retrieval (IR) systems capable of interpreting dense clinical information and delivering precise results.

In recent years, there has been a significant rise in the adoption of deep learning (DL) based methods for processing and extracting information from natural language text, both in the general and clinical domains. These DL methods have consistently outperformed traditional machine learning (ML) approaches across various natural language processing (NLP) tasks, including open-domain applications as well as those within the clinical domain \citep{PMID:31794016}. A pivotal moment in this evolution was the inception of transformer-based architectures\citep{vaswani2023attentionneed} , such as BERT \citep{devlin2019bertpretrainingdeepbidirectional} and its variants, which have revolutionized both the NLP and IR fields. Transformers introduced the ability to capture long-range dependencies and contextual relationships in text more effectively than previous approaches, also allowing for transfer learning and domain adaptation, leading to state-of-the-art performance across a wide range of tasks.

While NLP and DL has opened many doors to tackle Patient cohort retrieval, DPR has emerged as one of the most promising methods, given its wide application in domains such as RAGs \citep{lewis2021retrievalaugmentedgenerationknowledgeintensivenlp}, question answering (QA)\citep{Cao2011}, search engines, and more. Passage retrieval is a critical component of IR systems, where the goal is to identify and retrieve the most relevant segments of text or "passages" from a large corpus of documents based on a user’s query.

\begin{figure}[htp]
  \includegraphics[width=\columnwidth]{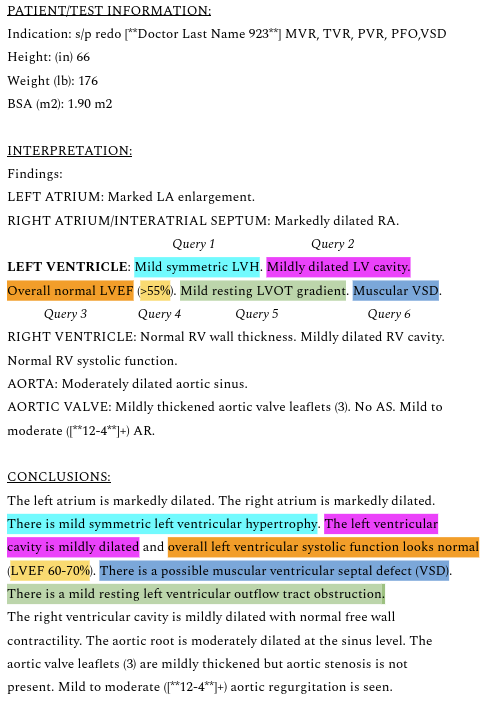}
  \caption{Echocardiographic report with queries highlighted in the same color in both the observations and summary sections.}
  \label{fig:report}
\end{figure}

Information retrieval (IR) in the medical domain poses unique challenges due to its inherent complexity. Medical terminology is rich and specialized, with abbreviations and technical jargon that vary significantly between contexts. %Many medical terms exhibit polysemy, where the meaning depends heavily on the surrounding text, adding to the ambiguity.%
Furthermore, the lack of standardization in EHRs, clinical notes, and medical prescriptions complicates efforts to extract structured and relevant data. Unlike standard QA tasks, medical texts like EHRs or clinical reports are often dense and information-rich, containing multiple conditions with varying severities described within the same sentence or passage as shown in Figure~\ref{fig:report}. Medical IR systems also require exceptionally high precision, as errors in retrieving relevant information can directly impact clinical decisions. 

The task of patient cohort retrieval (CR) in Echocardiography involves searching through a large set of EHR documents to identify groups of patients most relevant to a given query. For instance, one might need to retrieve patients with "left ventricular ejection fraction (LVEF) < 40\%" for studying heart failure. In this scenario, a CR system is expected to return a list of patients who have an LVEF measurement below 40\% at any point in their echocardiography records. This type of query-to-document matching requires an understanding of general vocabulary (e.g., "<40\%" implies patients with measurements less than 40\%), domain-specific vocabulary (e.g., LVEF might also be referred to as "ejection fraction" or simply "EF" in the notes). Specifically, we make the following main research contributions:
\begin{itemize}
    \item Identifying and preparing a suitable dataset tailored to our use case
    \item Designing evaluation sets that reflect real-world clinical scenarios
    \item We propose modifications to the commonly used loss for DPR, Multiple Negatives Ranking Loss to better suit cohort retrieval tasks
    \item Developing a specialized custom DPR trained efficient Embedding model for this task
    \item We compare different approaches in the task of Cohort retrieval along with ours
\end{itemize}

\section{Related Work}

The field of cohort retrieval has grown increasingly sophisticated over the years, paralleling advancements in NLP and IR. Early approaches primarily relied on rule-based methods to identify cohorts based on phenotypic characteristics \citep{Shivade2013ARO}. Regex-based techniques were employed to extract quantitative patterns, such as left ventricular ejection fraction (LVEF) values, from echocardiography reports, as seen in the work of \citealp{Wagholikar2018ExtractionOE}. Structured Boolean queries have also been used to define precise rules for filtering cohorts, enabling automated word-based retrieval methods \citep{Chamberlin2019EvaluationOP}. While these methods provided initial solutions using rules and heuristics, they were often limited by their inability to handle the lexical diversity and complexity of clinical language.

Some approaches focused on query extraction and expansion, aiming to better capture the intent behind clinical queries and rank patient visits associated with specific cohorts \citep{Goodwin2018LearningRM}. Other methods like CREATE \citep{Liu2019CREATECR} or EliIE \citep{Kang2017EliIEAO} used Common data models (CDM) like OMOP (Observational Medical Outcomes Partnership), a standardized framework for organizing healthcare data from various sources, likewise \citealp{Chang2022CohortIF} utilized SNOMED CT (Systematized Nomenclature of Medicine Clinical Terms), a collection of medical terms used in clinical documentation and reporting to convert unstructured EHRs to standardized data representation for easy structured data filtering. 

Entity extraction has also been a dominant approach, which involved identifying and classifying relevant medical concepts from clinical text. For instance, \citealp{Kusa2023EffectiveMO} used named entity recognition (NER) and negation detection to extract medical entities from patient descriptions and trial eligibility sections, leveraging a Transformer-based architecture. Another approach trained attention-based bidirectional Long Short-Term Memory (Att-BiLSTM) models to extract variables from COVID-19 trial eligibility criteria \citep{Liu2020AttentionBasedLN}. Likewise, \citealp{Nye2020UnderstandingCT} employed Conditional Random Fields (CRF) layered on top of BERT + biLSTM model for entity recognition tasks. However, these systems face challenges such as low recall, limited generalizability, and reliance on extensive manual annotations. LLMs also have shown enormous promise for NER in a zero shot manner as explored by \citealp{monajatipoor2024llmsbiomedicinestudyclinical}, they are not as efficient as the DPR approach, also the restrictions that come with dataset usage prevents us from using Closed SOTA LLMs APIs.

Other methods adopted IR frameworks for end-to-end cohort retrieval. For example, embedding-based method \citep{Glicksberg2018AutomatedDC} used word2vec to learn to encode both patient data and queries into a shared representation space for retrieval tasks. Similarly, models based on pre-trained architectures like BERT have been fine-tuned to encode clinical trial information into sentence embeddings for matching queries with relevant patient data \citep{zhang2020deepenrollpatienttrialmatchingdeep}. Another strategy trained BERT model to classify query-report pairs as "relevant" or "non-relevant," optimizing cohort retrieval through supervised learning \citep{Soni2020PatientCR}.
%Approaches like Multiple Instance Learning (MIL) further extend this paradigm, organizing training data into "bags" of instances and learning to classify the relevance of the entire bag \citep{Dai2020CohortSF}.%

DPR gained significant attention following its introduction by \citep{karpukhin-etal-2020-dense}, which demonstrated its effectiveness in efficiently retrieving relevant information. DPR employs a dual-encoder architecture: one encoder generates embeddings for queries, and the other encodes passages. These embeddings are compared using a similarity metric, such as dot product, to retrieve the most relevant passages. Its potential has since been explored in biomedical applications, as highlighted by studies \citealp{Gupta2023TopKR}, \citealp{rosso-mateus-etal-2020-deep}, and \citealp{Luo2022ImprovingBI}. Despite its wide adoption in mainstream domains, we have not seen enough public literature on the topic of DPR for cohort retrieval tasks.

\begin{figure*}[htp]
  %\centering
  \includegraphics[scale=0.4]{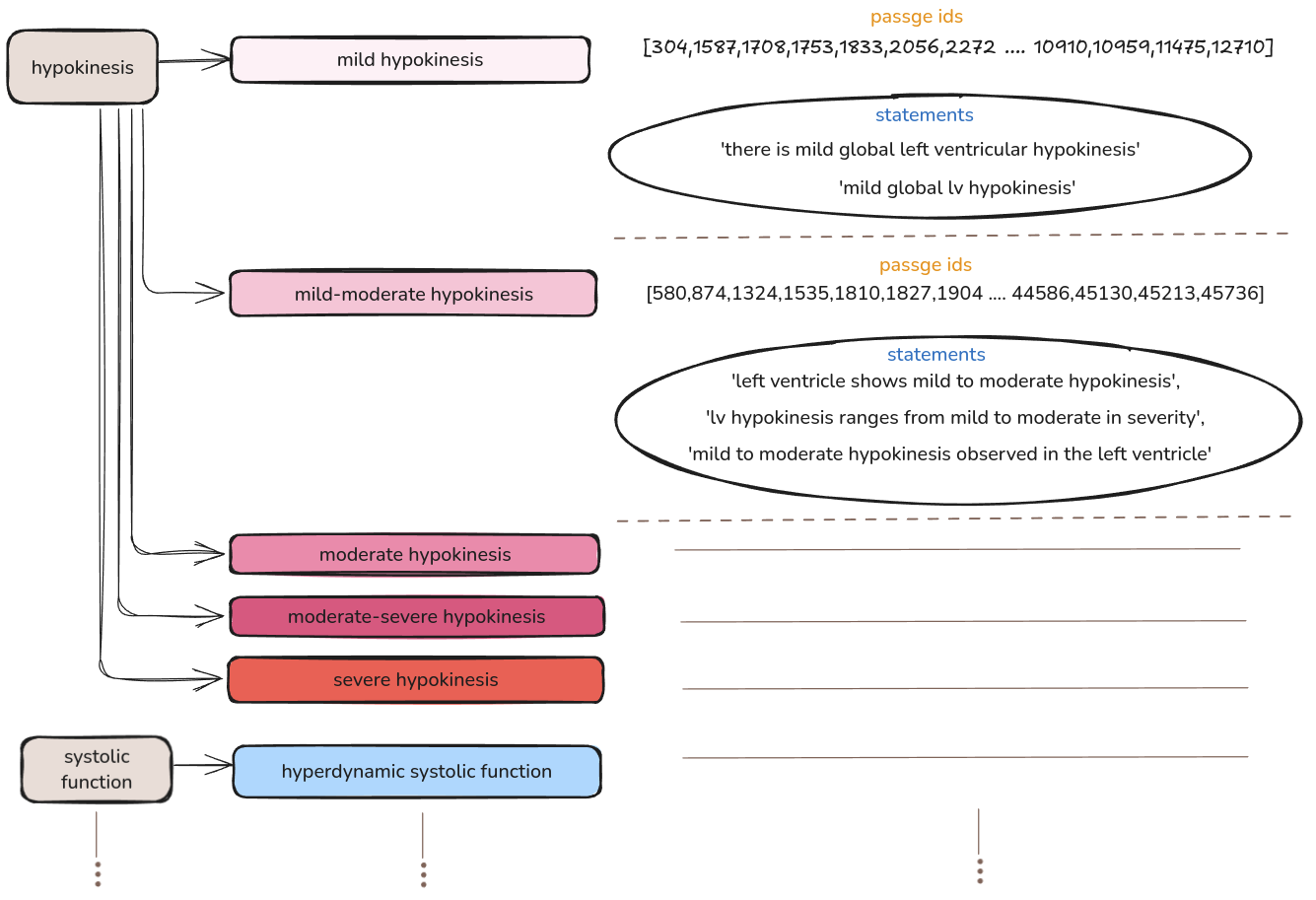}
  \caption{Systematic mapping of conditions, subcategories, statements, and associated passage IDs}
  \label{fig:conditions}
\end{figure*}

To successfully learn knowledge of a specialized domain like echocardiography in our case, techniques like pretraining have been readily deployed. Pretraining BERT on medical domain-specific corpora as shown by \citealp{Gu2020DomainSpecificLM}, \citealp{9628032} significantly enhances its performance in healthcare applications. This specialized pre-training enables the model to grasp the unique terminology, abbreviations, and contextual nuances prevalent in medical texts, leading to more accurate interpretations. Pretraining often lies under unsupervised learning domain and hence can be fed enormous data without the need of any explicit labels. Consequently, downstream tasks such as NER, relation extraction, and clinical document classification or passage retrieval benefit from pre-training improving their precision and recall.

\section{Materials and Methods}

\subsection{Dataset}
\label{subsection:dataset_name}

In this work, we utilized the EchoNotes Structured Database derived from the MIMIC-III (ECHO-NOTE2NUM) \citep{Kwak2024}, a comprehensive dataset comprising 43,472 echocardiography reports from the Beth Israel Deaconess Medical Center's intensive care unit, collected between 2001 and 2012. It provides a comprehensive set of labels that quantify various aspects of cardiac structure and function. These labels are derived from echocardiography reports and are systematically categorized to facilitate detailed analysis. While ECHO-NOTE2NUM covers in total 19 such labels across multiple anatomies, for Left Ventricle it only has systolic and diastolic function, LV cavity and wall thickness and motion, we do our own annotations to cover a wide variety on top of this, like \textit{gradient, cardiac index, cardiomyopathy, hypokinesis, mass/thrombus, dyssynchrony, VSD, LVH, aneurysm, PCWP, LVEF.}

Due to confidentiality, we will not be able to make these annotations public. Our annotations were created using automated heuristics and rules, with manual supervision to ensure accuracy.  A local Large Langauage Model (LLM), Meta-LLama-8b \citep{Dubey2024LLaMA3}, was also used to speed up the process and cluster variations of the same condition and function. 

Each report consists of patient information, finding section, and conclusion section. All reports have been de-identified of any PHI information. The Findings section provides a detailed, structured analysis of the heart's various components. It includes measurements and observations for each anatomical feature assessed during the echocardiographic examination, it documents specific data points, including chamber sizes, wall thicknesses, valve functions, and any detected abnormalities. This comprehensive information serves as the foundation for clinical assessment and decision-making.

The Summary section distills the detailed findings into a concise overview, highlighting the most significant observations and their clinical implications. It synthesizes the detailed information from the Findings section, providing a brief account of the patient's cardiac status. This section serves as a quick reference for healthcare providers, facilitating efficient communication and understanding of the patient's condition. 

\subsection{Dataset Preparation}

The Findings section is composed of individual conditions and observations, each documented in separate sentences, making it straightforward to parse and simplify into queries. Each statement in the Findings section is explicit and structured, allowing us to generate a set of distinct, condition-specific queries as shown in Figure~\ref{fig:report}.

%The Summary section, in contrast, provides a high-level overview and may paraphrase, condense, or omit some redundant details mentioned in the Findings section. While it may not always include the exact wording found in the Findings, the Summary conveys an overall interpretation that aligns with the detailed observations.
%
For this work, Due to annotation constraints we relied only on Left Ventricle section within the Findings to extract relevant query terms. Only reports containing an LV section were included in the dataset, and those without it were excluded. We parsed, split, and collated all LV-related statements from the findings section across the reports around 130k, ultimately identifying approximately 400 unique statements (queries) to use as the basis for cohort retrieval tasks. For passages, we used the Summary sections, selecting a total of 42,996 summaries corresponding to reports with LV observations. 

We devised a structured approach to categorize and organize these statements, assigning each statement to one or more specific clinical conditions. 
For example, a statement such as 
\textit{"Left ventricular wall thickness, cavity size, and systolic function are normal"} was categorized under the conditions \textit{“wall thickness”, “cavity size”, and “systolic function”}. This process was applied across all statements in the dataset, resulting in a comprehensive mapping of each statement to relevant cardiac conditions.

During this categorization, we observed that certain conditions, like hypokinesis, exhibited multiple subcategories based on severity or extent, such as mild, mild-moderate, severe, and global hypokinesis. To account for these nuances, we further clustered the individual condition statements into distinct subcategories as shown in Figure~\ref{fig:conditions}.

Additionally, we identified multiple variations of statements describing the same subcategories. For example, under hyperdynamic systolic function, we encountered variations such as:
\textit{
\begin{itemize}
    \item "Left ventricular systolic function is hyperdynamic"
    \item "Hyperdynamic left ventricular systolic function"
    \item "Hyperdynamic LV systolic function"  
\end{itemize}
}
These variations were grouped together under the concept of hyperdynamic systolic function to ensure variation in queries.
By maintaining an index of each statement’s occurrence within the corresponding echocardiography reports, we were able to construct an inverted index. This index maps each categorized statement to the list of relevant passages in which it appears, allowing us to link queries (statements) directly to passages as shown in Figure~\ref{fig:conditions}. This structured organization of statements into categorized conditions, subcategories, and their corresponding passages facilitates condition-based cohort retrieval dataset.

\subsection{Retrieval Tasks and Evaluation Sets}
To comprehensively evaluate our system's performance across diverse retrieval tasks, we employed several techniques to enhance our evaluation set that challenge the model's ability to generalize, handle out-of-distribution (OOD) conditions, understand quantitative information, and interpret paraphrased text effectively.

\subsubsection{Held-Out Set for Evaluation}
\label{subsubsection:heldout}
We reserved a held-out evaluation set, comprising an entirely different set of passages (around 9k passages) corresponding to the same set of queries used in the training set. Additionally, we excluded queries for certain subcategories (four in total) of specific conditions from the training set, ensuring that these were only present in the evaluation set. 
%This approach allows us to assess the model's generalization capability on conditions not encountered during training.
Specifically, we excluded the following subcategories:
\begin{itemize}
    \item “Mild-Moderate Hypokinesis”
    \item “Apical Intracavitary Gradient” and “No Apical Intracavitary Gradient”
    \item “Mural Thrombus”
\end{itemize}
This held-out set, consisting of unique query-passage pairs, provides a robust evaluation framework for testing the model's ability to generalize to new variations of known conditions.

\subsubsection{Out-of-Distribution (OOD) Condition Evaluation} 
To assess the model’s performance on OOD tasks — rare conditions like \textbf{cor triatriatum, Ebstein’s anomaly, and LV noncompaction} were chosen because they are clinically significant but infrequently encountered, making them ideal for testing the model’s ability to handle OOD problems. These conditions are only present in single-digit occurrences within our dataset. Any trace of these conditions was entirely removed from the DPR training process. During OOD evaluation the passages of OOD conditions were added to the held out set as mentioned in section \ref{subsubsection:heldout} and evaluated using relevant queries for the rare conditions.

\subsubsection{Numerical Value Retrieval Tasks}  
\label{subsubsection:numeric}
In the context of quantitative retrieval tasks, particularly within medical reports like echocardiography notes, precise measurements such as Left Ventricular Ejection Fraction (LVEF) are frequently expressed with numerical ranges or signs. To effectively handle these numerical conditions, we employed heuristic parsing methods to extract LVEF values, signs, and ranges from the passages, creating a temporary database that links each LVEF mention to its relevant passage. This approach aligns with the methodologies proposed by \citealp{almasian-etal-2024-numbers}, who introduced quantity-aware ranking techniques designed to incorporate numerical information into retrieval systems, addressing queries with numerical conditions such as =, >, < signs. Building upon this framework, we automatically generated evaluation queries with varying paraphrasing, signs, and numeric values to assess retrieval performance on numerical data. By referencing our temporary database, we identified relevant positive and negative passages for each query, forming an LVEF-specific evaluation set of 100 LVEF specific queries relevant to the same held out passage set in section \ref{subsubsection:heldout}.

\subsubsection{Paraphrased Query Set}  

To further evaluate the model’s ability to handle linguistic variation, we augmented the original query set by generating paraphrased versions using the Meta-llama3-8b model. We used prompts designed to preserve medical meaning while introducing linguistic diversity, ensuring that numerical values and clinical terms were retained in context. We encourage the generative model to use double negations, antonyms, synonyms, and simplified descriptions of conditions. This approach creates multiple variations of each query, expanding the diversity of expressions. A total of 60 new paraphrased queries were created, for these paraphrased queries, the relevant passages remain the same in held out set as for the original queries.

\begin{table}[htbp]
    \begin{tabular}{lcccc}
        \hline
        \textbf{Statistic} & \textbf{Mean} & \textbf{Std} & \textbf{Min} & \textbf{Max} \\
        \hline
        Passage Length & 239 & 88 & 10 & 868 \\
        Query Length   & 11  & 5  & 3  & 30  \\
        \hline
    \end{tabular}
    \caption{Token Statistics for Passages and Queries}
    \label{tab:desc_stats}
\end{table}

\subsection{Evaluation Metrics}

%\begin{strip}
\begin{figure*}
\centering
\begin{equation}
    \mathcal{L} = -\frac{1}{N} \sum_{i=1}^N \log \frac{\exp(\text{sim}(a_i, p_i))}{\exp(\text{sim}(a_i, p_i)) +\exp(\text{sim}(a_i, n_i)) + \sum_{j \not= i}\exp(\text{sim}(a_i, p_j)) + \sum_{j \not= i}\exp(\text{sim}(a_i, n_j))}
    \label{eq:loss_function}
\end{equation}
\end{figure*}
%\end{strip}

Typically, Traditional IR systems are evaluated using metrics that assess both relevance and ranking quality. But in cohort retrieval, we are only interested in relevance and not rank-aware metrics. We evaluated our model's performance using metrics tailored to our specific requirements, given that the number of relevant passages varies across queries. Our objective is to ensure that a relevant passage, if available, is retrieved, which we consider a success. For this reason, we use 

\begin{itemize}
  \item \textbf{Precision@10} and \textbf{Precision@100}: These metrics measure the proportion of relevant passages among the top 10 and top 100 retrieved passages, respectively.
  \item \textbf{R-Precision}: This metric calculates the fraction of relevant passages among the top R retrieved passages. Specifically, R-Precision is defined as:
    \begin{equation}
      \label{eq:example}
      R\mathrm{-}Precision = \frac{n\_relevant}{R}
    \end{equation}
    where:
    \begin{itemize}
        \item n\_relevant indicates the number of relevant passages retrieved within the top \textit{R} results
        \item \textit{R} is the total count of relevant passages for the query
    \end{itemize}
\end{itemize}

\section{Experiment Details}

In our experiments, For the DPR task, we trained an embedding model using msmarco-bert-base-dot-v5 model as architecture. This model is a specialized variant of the BERT architecture, originally fine-tuned for semantic search tasks, and built on the BERT-base architecture with 12 transformer layers and 12 attention heads, totaling approximately 110 million parameters. The model generates 768-dimensional dense vector embeddings for sentences and paragraphs, making it highly effective for semantic similarity computations. We trained this model in a bi-encoder fashion with a single encoder shared for both query and passage embedding generation rather than using separate encoders.

We fixed the model’s maximum sequence length by analyzing the token distribution for reports and queries in our dataset as shown in Table \ref{tab:desc_stats} to be 300, anything less than that was padded and anything more than that was truncated. A training batch size of 64 was used, and mean pooling was applied to aggregate word embeddings into sentence-level embeddings. The Warm-up Linear learning rate scheduler was employed, with a learning rate of \(2 \times 10^{-5}\) and 1000 warmup steps. The model was trained for 200,000 steps (approximately 31,250 batches) using the AdamW optimizer. We have utilized the default hyperparameters provided by the Sentence Transformers library's training script \citep{reimers2019sentencebertsentenceembeddingsusing}. The model was trained from scratch without any prior pretrained weights. To optimize the model for retrieval, Multiple Negatives Ranking Loss was selected as the loss function. 

While Multiple Negatives Ranking (MNR) Loss \citep{henderson2017efficientnaturallanguageresponse} is generally designed for (query, positive) pairs, we also supplied a hard negative sample with each pair. The loss function works by minimizing the distance between the anchor \(a_i\) embedding and its positive embedding counterpart \(p_i\) while maximizing the distance between the anchor embedding and the negative embedding \(n_i\) while also using other positive and negatives of other pairs as negative samples, also known as in-batch negative sampling as shown in Equation~\ref{eq:loss_function}.
Where:
\begin{itemize}
    \item \(a_i\): The query or Anchor embedding for the \textit{\(i_{th}\)} pair.
    \item \(p_i\): The positive passage embedding paired with the query \(a_i\).
    \item \(n_i\): Embedding of Explicitly supplied hard negative for each pair with the query \(a_i\).
    \item \(p_j\): Embeddings of positive passages of other pairs in the batch.
    \item \(n_j\): Embeddings of negative passages of other pairs in the batch.
    \item sim(x,y): The similarity function (e.g., cosine similarity) between embeddings x and y.
    \item N: Total number of pairs in the batch.
 
\end{itemize}
Since MNR Loss utilizes other pairs' positive and negative passages as in-batch negatives, it is not directly suitable for our task. In our case, many queries are associated with multiple relevant passages, leading to a high likelihood that another pair's positive or negative passage could actually be a valid positive passage for the current anchor query. This would result in misleading gradients and hinder the model's learning.

To address this, we modified the loss computation by setting the cosine similarity scores of any positive passages (other than the designated anchor-positive pair) in the batch to $ -\infty $. This adjustment ensures that these additional positive passages are excluded from contributing to the loss calculation.
%By doing so, we prevent these passages from being treated as negatives and eliminate potential confusion for the model, allowing it to focus on the correct anchor-positive relationships during training.

For our v0 model variant(echo-retriever-model v0), we trained solely on the dataset’s training split without any data augmentations. We created a dataloader which yielded triplets where each triplet consisted of a query, a positive passage, and a negative passage. To construct a triplet, we first selected a condition from the dataset and then randomly sampled two distinct subcategories within that condition. One of these subcategories was assigned as the positive example and the other as the negative. From the positive subcategory, a query was randomly chosen from the list of associated statements. Similarly, a passage related to the positive subcategory was selected as the positive passage, while a passage associated with the negative subcategory was chosen as the negative passage.
For conditions that did not have multiple subcategories, we used a neutral passage from a different condition as the negative example. We ensured that the negative passage was not relevant to the query or positive passage, maintaining the integrity of the triplet structure. Through this approach of hard negative sampling, we aimed to create challenging training pairs that would help the model learn fine-grained distinctions between similar but non-identical statements. The best model checkpoint was selected based on performance on a separate, hidden evaluation set using precision @ 10 as metric. 

\begin{table*}
  \centering
  \begin{tabular}{ll}
    \hline
    \textbf{Category} & \textbf{Expression Examples} \\
    \hline
    \multirow{3}{*}{For specific values} & \verb|"LVEF {sign} {mid_value}%"| \\
                                          & \verb|"EF {sign} {mid_value}%"| \\
                                          & \verb|"ejection fraction {sign}{mid_value}%"| \\
    \hline
    \multirow{3}{*}{For ranges} & \verb|"LVEF between {min_value}-{max_value}"| \\
                                  & \verb|"LVEF measured at {min_value}% to {max_value}%"| \\
                                  & \verb|"Expected LVEF in range of {min_value}% to {max_value}%"| \\
    \hline
  \end{tabular}
  \caption{Expressions for Specific Values and Ranges for LVEF}
  \label{tab:expressions}
\end{table*}

For v1 model variant(echo-retriever-model v1), we extended the v0 model variant training framework by augmenting the training data with triplets specifically designed for quantitative retrieval tasks and for tackling paraphrased queries tasks. To achieve this, we utilized the same temporary LVEF database that was created as mentioned in the section \ref{subsubsection:numeric}. Queries for LVEF were created using a list of templated statements, with placeholders that were populated in real time during training by randomly sampling numeric values, ranges, and signs. Sample query templates as shown in Table \ref{tab:expressions}

This setup enabled a diverse range of queries involving different signs (>, <, =) , numeric values and query formats supporting robust LVEF retrieval training. To improve generalization to paraphrased queries, statements within each subcategory were further augmented using Meta-llama3-8b LLM model to create paraphrased versions. It was ensured that the training set of LVEF task and paraphrasing queries task is exclusive of the one created for evaluation, warranting no leakage.

For the v2 model variant(echo-retriever-model v2), we performed Masked language modeling (MLM) \citep{devlin2019bertpretrainingdeepbidirectional} pre-training to deepen the model's understanding of the underlying corpus before applying the same training framework used for the v1 model variant. The intuition was to improve the embedding model's performance on OOD tasks without requiring direct supervised learning. 

MLM pre-training helps the model learn robust contextual embeddings by predicting masked tokens. For MLM training, we applied whole word masking technique with a masking probability of 0.15 to learn contextual information more effectively. The model was trained with a batch size of 64 for a total of 71,600 batches, using the AdamW optimizer with a learning rate of $5 \times 10^{-5}$. We utilized the complete dataset for this unsupervised pre-training phase, which aimed to enhance the model’s comprehension of domain-specific language patterns, preparing it for downstream retrieval tasks in the medical domain. In this pretraining variant, the model excludes the pooling layer typically used for generating sentence embeddings, as it focuses solely on learning token-level representations. These hyperparameters were selected based on the training script from the sentence-transformers library. After MLM pre-training was completed, the model weights were used as a starting point for fine-tuning the v2 variant.

\section{Results}

\begin{table*}[htp]
  \centering
  \renewcommand{\arraystretch}{0.9} % Adjusts row height
  \setlength{\tabcolsep}{4pt} % Adjusts column separation
  \begin{adjustbox}{width=\textwidth}
  \begin{tabular}{lccccccccccccccc}
    \hline
    \multirow{2}{*}{Models} & \multicolumn{3}{c}{Held Out Set} & \multicolumn{3}{c}{Paraphrased Queries} & \multicolumn{3}{c}{Numerical Tasks} & \multicolumn{3}{c}{OOD} \\
    \cmidrule(lr){2-4} \cmidrule(lr){5-7} \cmidrule(lr){8-10} \cmidrule(lr){11-13} \cmidrule(lr){14-16}
    & P@10 & P@100 & R-Prec & P@10 & P@100 & R-Prec & P@10 & P@100 & R-Prec & P@10 & P@100 & R-Prec \\
    \hline
    BM25 & 0.49 & 0.43 & 0.37 & 0.28 & 0.26 & 0.24 & 0.10 & 0.14 & 0.18 & \bfseries 0.48 & \bfseries 0.48 & \bfseries 0.48 \\
    ClinicalBert & 0.24 & 0.19 & 0.18 & 0.08 & 0.07 & 0.08 & 0.16 & 0.16 & 0.18 & 0.36 & 0.36 & 0.36 \\
    jina-embeddings-v2-base-en & 0.34 & 0.30 & 0.27 & 0.26 & 0.22 & 0.19 & 0.21 & 0.21 & 0.20 & 0.16 & 0.16 & 0.16 \\
    echo-retriever-model v0 & \bfseries 0.85 & \bfseries 0.82 & \bfseries 0.78 & 0.66 & \bfseries 0.63 & \bfseries 0.60 & 0.25 & 0.25 & 0.18 & 0.06 & 0.06 & 0.06 \\
    echo-retriever-model v1 & 0.81 & 0.76 & 0.75 & 0.63 & 0.59 & 0.56 & 0.98 & 0.98 & 0.98 & 0.06 & 0.06 & 0.06 \\
    \bfseries echo-retriever-model v2 & 0.82 & 0.77 & 0.76 & \bfseries 0.68 & 0.62 & 0.59 & \bfseries 0.99 & \bfseries 0.98 & \bfseries 0.98 & 0.00 & 0.00 & 0.00 \\
    \hline
  \end{tabular}
  \end{adjustbox}
  \caption{Model Performance Across Various Tasks}
  \label{tab:model_performance}
\end{table*}

The evaluation results for traditional models and the best off-the-shelf state-of-the-art (SOTA) models, in comparison with our fine-tuned embedding based retrieval model across various retrieval tasks are shown in the Table~\ref{tab:model_performance}.

BM25 \citep{10.1145/299917.299920} serves as a strong baseline and significantly outperforms other DL based off-the-shelf methods BERT models. Although BM25 is not optimized for DPR, its term-based ranking approach provides competitive performance, particularly when exact or near-exact term matches are common, which reflects its robustness as a  traditional IR baseline that remains competitive for term-based retrieval tasks.

Given that there is no SOTA DPR model that has been trained specifically for the echocardiography domain, we included ClinicalBERT \citep{huang2020clinicalbertmodelingclinicalnotes} as the closest available option. ClinicalBERT was pre-trained on a large, multi center corpus of EHRs, encompassing over 1.2 billion words from 3 million patient records, covering diverse diseases and medical conditions. However, despite its extensive pre-training on clinical language, ClinicalBERT struggles across all tasks, performing poorly compared to other off-the-shelf methods. This suggests that although ClinicalBERT is optimized for general clinical language understanding, it lacks the specialized training required for effective passage retrieval, especially for tasks involving nuanced and domain-specific queries.

Jina-embeddings-v2-base-en \citep{guenther2024jinaembeddings28192token}, another off-the-shelf model, is a high-capacity English embedding model with support for long sequence lengths (up to 8192 tokens). It is based on the jina-bert-v2-base-en backbone, which was pretrained on the C4 dataset and further fine-tuned on over 400 million sentence pairs, including hard negatives from diverse domains. This model is designed for various use cases, including long document retrieval, semantic similarity, and text ranking. However, while Jina-embeddings-v2 offers a strong general-purpose model performance for semantic search tasks, it falls short in domain-specific retrieval tasks like echocardiography cohort retrieval, where domain adaptation and specialized training are crucial. However it performs better than ClinicalBERT, reinforcing the deductions made by \citealp{Excoffier2024GeneralistEM}.

The Echo-retriever-v0 variant, trained specifically on the EchoNotes dataset \ref{subsection:dataset_name}, demonstrates marginally superior performance on the held-out set compared to other model variants. However, it significantly under performs on the numerical retrieval tasks, revealing the need for specialized data and training strategies to handle numerical values effectively which are eventually solved in Echo-retriever-v1 variant.

All models trained on the EchoNotes dataset perform poorly on OOD tasks, highlighting a critical weakness in generalizing to rare or previously unseen conditions. Although OOD conditions were included in the MLM pre-training stage, this pretraining did not transfer effectively to the fine-tuned model Echo-retriever-v2 variant. Interestingly, BM25 achieves the best performance on OOD tasks by identifying passages with the correct conditions, yet it struggles to differentiate between varying levels of severity, which further illustrates its limitations in more granular retrieval scenarios. 

%The unexpected drop in performance for v1 and v2 variants in Held out Set can be attributed to inclusion of more tasks and augmentations during training which %added noise rather than generalization. While there is a small drop in performance but the trade offs were acceptable given the significant increase in other %tasks. MLM Pretraining used in v2 variant also did not have any prominent upside on any tasks and was marginally better than v1 variant.

In summary, while our domain-specific fine-tuned models outperform off-the-shelf SOTA models and traditional baselines like BM25 in most retrieval tasks, they still face challenges in OOD generalization. These findings emphasizes the importance of specialized data, task-specific training, and potential enhancements in model architecture to improve retrieval accuracy in medical domain applications.
If looking at the overall picture, we can call that the model v2 variant as the best one as it performs decently well on all tasks other than OOD.

\section{Conclusions}
This work addresses the challenges of patient cohort retrieval in echocardiography by leveraging DPR techniques. We showcased how to transform an EHR dataset into a passage retrieval task which can interpreted as Cohort retrieval problem. We also demonstrated the superiority of our custom DPR trained embedding model over other off the shelf SOTA models. Additionally, we proposed modifications to the Multiple Negatives Ranking Loss to better suit cohort retrieval tasks. Our findings emphasize the importance of task-specific training and evaluations. While this work focuses on echocardiography-specific conditions, the methodology and insights presented are adaptable to other medical domains with appropriate datasets and evaluation tasks. To the best of our knowledge, this is the first work to explore the application of DPR for patient cohort retrieval in the echocardiography domain, marking a significant step forward in domain-specific information retrieval.

\section*{Limitations}

This study has several limitations, both in terms of the dataset and the proposed models. First, our comparisons were restricted to embedding-based models, specifically those based on sentence transformers. While incorporating rule-based NER models or LLM-based NER off the shelf models would have provided a more comprehensive benchmark, doing so would have significantly broadened the scope of this paper.
Moreover, despite the impressive performance of LLMs in retrieval tasks within context-rich domains under zero-shot conditions, their applicability to patient cohort retrieval remains constrained. Specifically, encoding cohort database indices within the input context length of LLMs would be infeasible, limiting their effectiveness for this task. We also did not try to determine the impact of model parameters or different architectures of S-Bert models on the performance. Since as we have shown our best model struggles with OOD task, future research should also investigate methods for training such models for incorporating OOD knowledge to improve their generalizability. For the dataset side, we limited ourselves to Left ventricle anatomy only due to annotation concerns, additional anatomies would have helped in unlocking new dimensions in retrieval complexities.

\bibliography{anthology,custom}
\end{document}